\documentstyle[sprocl]{article}

\def\o{\over}
\def\Ar{\rightarrow}
\def\bar{\overline}

\def\a{\alpha}

\def\n{\nu}
\def\m{\mu}

\def\th{\theta}

\def\bar{\overline}
\def\l{\lambda}

\def\eV{{\rm eV}}

\def\be{\begin{equation}}
\def\ee{\end{equation}}
\def\bea{\begin{eqnarray}}
\def\eea{\end{eqnarray}}

\begin{document}

\title{Generation of Neutrino Masses and  Mixings in Gauge Theories}

\author{Morimitsu Tanimoto}

\address{Ehime University, Matsuyama, JAPAN: tanimoto@edsrev.ed.ehime-u.ac.jp} 


\maketitle\abstracts{ I review  models which present large flavor  mixings
of the lepton sector based on the gauge theory. (Invited talk at WIN99)}

\section{Introduction}
Recent experimental data of neutrinos make big impact on the neutrino
masses and their mixings.
Most exciting one is the results at Super-Kamiokande on the atmospheric
neutrinos,
 which indicate the large neutrino flavor oscillation of $\n_\m\Ar \n_x$
\cite{SKam}.
 Solar neutrino data also provide the evidence of the neutrino oscillation,
however
 this problem still uncertain \cite{BKS}. 

What can we learn from these results?
We want to get clues for  origins of neutrino masses and neutrino flavor
mixings.
In particular, we want to underatand why the neutrino mixing is large
compared with the quark sector.
Now we should discuss these problems in connection with the quark sector.

 \section{Phenomenological Aspect of Neutrino Masses and Mixings} 
 Our starting point as to the neutrino mixing is
  the large $\nu_\m \Ar \nu_\tau$ oscillation of the atmospheric neutrino
oscillation with 
 $\Delta m^2_{\rm atm}=  (1\sim 6)\times  10^{-3} \eV^2$ and 
 $\sin^2 2\th_{\rm atm} \geq 0.9$
 which are derived from the recent data of the atmospheric neutrino deficit
at Super-Kamiokande
   \cite{SKam}. 
   In the solar neutrino problem \cite{BKS},
 there are three solutions:
  the MSW small angle solution,  the MSW large angle solution and the
vacuum solution.
  These mass difference scales are much smaller than the atmospheric one.
  Once we put $\Delta m_{\rm atm}^2=\Delta m_{32}^2$ and $\Delta
m_{\odot}^2=\Delta m_{21}^2$, 
 there are two typical mass patterns: $m_3\gg m_2\gg m_1$ and
 $m_3\simeq m_2\simeq m_1$.
 
 The neutrino mixing is  defined  as
  $\n_\a=U_{\a i}  \n_i$,      
   where $\a$ denotes the flavor $e,\m,\tau$  and  $i$ denotes mass
eigenvalues $1,2,3$.
  Now we have two typical mixing patterns:
  \begin{equation}
   U_{\rm MNS} = \left (\matrix{ 1 & U_{e2} & U_{e3} \cr
                          U_{\m 1} & {1\o \sqrt{2}} & -{1\o \sqrt{2}} \cr
		 U_{\tau 1} & {1\o \sqrt{2}} & {1\o \sqrt{2}} \cr	} \right )     ,
		 \qquad
	\left (\matrix{ {1\o \sqrt{2}} & -{1\o \sqrt{2}} & U_{e3} \cr
                         {1\o 2} & {1\o 2} & -{1\o \sqrt{2}} \cr
		  {1\o 2} & {1\o 2} & {1\o \sqrt{2}} \cr	} \right )     ,
 \end{equation}
  \noindent  the first one is the single maximal mixing pattern, in which
   the solar neutrino deficit is explained by the small angle
   MSW solution,  and the other is the bi-maximal mixings pattern, in which 
  the solar neutrino deficit is explained
   by the just so solution.  In both case $U_{e3}$ is constrained by the
CHOOZ Data.
 %
introduced.  These
quarks and leptons
  
\section{Neutrino Masses and Mixings in the GUT}
The left handed neutrino masses are supposed to be at most ${\cal O}(1) \eV$.
In the case of Majorana neutrino, we know two classes of models 
which lead naturally to a small neutrino mass:
(i) models in which the seesaw mechanism works and (ii) those in which the
neutrino mass is 
induced by a radiative correction.  
The central idea of  models (i) supposes some higher symmetry 
which is  broken at an high energy scale. If this symmetry breaking takes place 
so that it allowes the right-handed  neutrino to have a mass, and a small
mass induced
for the left handed neutrino by the seesaw mechanism. 
 In the classes of model (ii) one introduces a scalar particle 
with a mass of the order of the electroweak (EW) energy scale which breaks
the lepton number
in the scalar sector. A left-handed neutrino mass is then induced by a
radiative 
correction from a scalar loop.
This model requires some new physics at the EW  scale. 

 
 {\bf SU(5) GUT:}
 \ In the standpoint of the quark-lepton unification, the charged lepton
mass matrix
  is connected with  the down quark one.
  The mixing following from the charged lepton mass matrix may be
considered to be
   small like  quarks in the hierarchical base.  However, this  expectation
   is not true if the mass matrix is non-Hermitian.
   In the SU(5), fermions belong {\bf 10} and {\bf 5*}:
 \begin{equation}    
  {\bf 10}:\  \chi_{ab} =u^c + Q + e^c , \qquad {\bf 5^*}:\  \psi^a  = 
d^{c1} + L, 
  \end{equation}          
\noindent where $Q$ and $L$ are SU(2) doublets of  quarks and leptons,
respectively. 
 The Yukawa couplings are given by $10_i 10_j 5_H$(up-quarks) and
 $5^*_i 10_j 5^*_H$(down-quarks and charge leptons)(i,j=1,2,3).
 Therefore we get $m_E= m_D^T$ at the GUT scale.
 
 It should be noticed that observed quark mass spectra and the CKM matrix
 only constrains the down quark mass matrix typically as follows:
 
 \begin{equation}         
   m_{\rm down} \sim  K_D \left (\matrix{    
                \l^4 & \l^3 & \l^4 \cr     
                  x  & \l^2 & \l^2 \cr    
                  y & z  &   1   \cr } \right )    \quad {\rm with} \quad
\l=0.22 \ .
      \end{equation}        
\noindent    
   Three unknown  $x,\ y,\ z$ are related to the left-handed charged lepton
mixing
    due to  $m_E= m_D^T$.
	The left(right)-handed down quark mixings are related to
   the right(left)-handed charged lepton mixings in the SU(5).
	Thefore, there is a source of the large flavor mixing
	in the charged lepton sector if  $z\simeq 1$ is derived from some models.
	This mechanism was nicely used by some authors \cite{SY,Alb,SU5}.
	
	In the case of  the SO(10) GUT, SO(10) breaking may lead to the large mixing
	in the charged lepton sector
	if an  asymmetric interaction in the family space exists \cite{Alb}.
  In conclusion, the $\n_\m-\n_\tau$ mixing could be maximal
	in some GUT models, which are consistent with the quark sector.

{\bf See-saw enhancement:}	
  \ The large mixing may come from the neutrino sector.
  It could be obtained in the see-saw mechanism as a consequence of a
certain structure of the
  right-handed Majorana mass matrix \cite{en1,en2}.  
  That is the so called see-saw enhancement of the neutrino
  mixing due to the cooperation between the Dirac and Majorana mass matrices.

  Mass matrix of light Majorana neutrinos $m_\n$ has the following form
  \begin{equation}          
   m_\n \simeq -m_D M_R^{-1} m_D^T \ ,      
  \end{equation}          
 \noindent  
 where $m_D$ is the neutrino Dirac mass matrix and $M_R$ is the Majorana
mass matrix
 of the right-handed neutrino components.
 Then, the lepton mixing matrix is \cite{en1}            
   $V_\ell = S_\ell^\dagger \cdot S_\n \cdot V_s$,       
  where $S_\ell$, $S_\n$ are transformations which diagonalize the Dirac mass
  matrices of charged leptons and neutrinos, respectively.
  The $V_s$ specifies the effect of the see-saw mechanism, i.e. the effects of
  the right-handed Majorana mass matrix.  It is determined by
   \begin{equation}          
   V_s^T m_{ss}V_s = diag(m_1,m_2,m_3) \ , \quad {\rm with}\quad 
        m_{ss} = -m_D^{diag} M_R^{-1} m_D^{diag} \ .       
  \end{equation}  
  \noindent
  In the case of two generations,  the mixing matrix $V_s$  is easily
investigated
   in terms of one angle $\th_s$.
   This angle could be maximal under the some conditions of parameters
   in the Dirac mass matrix and right handed Majorana mass matrix.
   That is the enhancement due to the see-saw mechanism.
   The rich structure of right-handed  Majorana mass matrix can lead to
    the maximal flavor mixing of neutrinos.

   {\bf Radiative neutrino mass:} 
   \ In the class of models in (ii),
   neutrino masses are induced from the radiative corrections.
   The typical one is the Zee model, in which charged gauge singlet scalar
   induces the neutrino mass \cite{Zee}.  In this model, 
   the previous predictions are consistent with LSND data and atmospheric
neutrino data.
   Then the soalr neutrino deficit was explained  by introducing the
   sterile neutrino.
   However new solution has been found in the framework of the Zee model.
   In the case of the inverse hierarchy $m_1\simeq m_2 \gg m_3$,
   the bi-maximal mixing, which is consistent with
   atmospheric and solar neutrinos,  is obtained \cite{ZeeNew}.
   
   The MSSM with R-parity violation can also give the neutrino masses and
mixings.
   The MSSM allowes renormalizable B and L violation.  
   The R-parity conservation forbids the B and L violation in the superpotential
    in order to avoid the proton decay.  However 
	the proton decay is avoided in the tree level if either of B or L
violating term vanishs.
	The simplest model is the bi-linar R-parity violating model with
$\epsilon_i H_u L_i$ 
	for the lepton-Higgs coupling \cite{KN}.
	This model provides the large mixing which is consistent with 
	atmospheric and solar neutrinos.
   
  \section{Flavor Symmetry and Large Mixings}  
 In the previous discussions, we assumed the family structure in the mass
matrices.
  However masses and mixings may suggest the some flavor symmetry.
  The simple flavor symmetry is U(1), which was discussed intensively by
  Ramond et al.\cite{Ramond}.
  In their model, they assumed
  (1) Fermions carry U(1) charge, (2) U(1) is spontaneously broken by
$<\theta>$, in which
  $\theta$ is the EW singlet with U(1) charge -1, and (3) Yukawa couplings
appear
   as effective operators
  \begin{equation}          
       h^D_{ij} Q_i \bar d_j H_d \left ({\theta\o \Lambda}\right )^{m_{ij}} + 
	   h^U_{ij} Q_i \bar u_j H_u \left ({\theta\o \Lambda}\right )^{n_{ij}}
+...\ ,     
  \end{equation} 
  \noindent
   where $<\theta>/ \Lambda=\lambda\simeq 0.22$.
   The powers ${m_{ij}}$ and  ${n_{ij}}$ are determined from the U(1) charges
   of  fermions in order that the  effective operators are  U(1) invariants.
   The U(1) charges of the fermions are fixed by the experimental data of the
   fermion masses and mixings. Then the model has  anomalous U(1). 
 
  Another typical flavor symmetry is $S_3$.
  The $S_{3L}\times S_{3R}$ symmetric mass matrix is so called 
 the democratic mass matrix \cite{Demo}, which  needs the large rotation in
order to move to
  the diagonal base.  In the quark sector, this large rotation is canceled
each other 
  between   down quarks and  up quarks.  
  However, the situation of the lepton sector is very different from
  the quark sector if the effective neutrino mass matrix $m_{LL}^\n$ is 
  far from the democratic one and the charged lepton one is still  the
democratic one. 
   Let us consider the neutrino mass matrices, which provide large mixings
\cite{FTY}:
The typical one  is 
 \begin{equation}
M_\n= {c_\n}
             \left ( \matrix{1 & 0 & 0 \cr
                            0 & 1 & 0 \cr
                            0 & 0 & 1 \cr
                                         } \right )
\quad +\left (\matrix{0 & \epsilon_\nu & 0 \cr
                 \epsilon_\nu  & 0 & 0 \cr
                  0 & 0 & \delta_\nu \cr} \right ) , \quad {\rm or}
		 \quad  
				  + \left ( \matrix{-\epsilon_\nu & 0 & 0 \cr
                 0  & \epsilon_\nu & 0 \cr
                  0 & 0 & \delta_\nu \cr} \right ) ,
  \end{equation}
 \noindent where the first term is the $S_{3L}$ symmetric effective mass
matrix and
 the second or the third is the $S_{3L}$ breaking one. 
 In the case of the first breaking matrix, 
 the large mixing of $(1-2)$ family sector is completely canceled
 between   the neutrino and the charged lepton sectors, however the large
mixing of the $(2-3)$ 
 family in the charged lepton sector is not canceled.
 So we have the large mixing in the lepton flavor mixing matrix.
If we adopt the latter  symmetry breaking matrix \cite{FTY,FX}, 
we obtain the lepton mixing matrix to be near bi-maximal
because the large mixings from the charged lepton mass matrix cannot be
canceled.
This case can accommodate the "just-so" scenario for the solar neutrino
problem due to neutrino 
oscillation in vacuum.
recent data
 
\section{Summary}
 Models depends on three phenomenological aspects.
 Is the mixing pattern  the single maximal mixing or bi-maximal mixing ?
 Is there sterile neutrino ?
 Are the neutrino masses degenerated or hirarchical ones?
 More precise solar neutrino data will answer the first and second questions.
 More precise atmospheric neutrino data and 
  the long baseline experiments can answer the second question.
 The double beta dacay experiments may answer the last question.
 We need more data in order to establish the model as well as more 
theoretical studies.

\end{document}